\documentclass[journal = jctcce, manuscript=article]{achemso}
\usepackage{ulem}

\newcommand{\bpsi}{\boldsymbol{\psi}}
\newcommand{\bpsibar}{\bar{\boldsymbol{\psi}}}

\newcommand{\al}{_{\alpha}}

\newcommand{\bd}{\boldsymbol{\textbf{d}}}

\newcommand{\bq}{\boldsymbol{q}}
\newcommand{\br}{\boldsymbol{\textbf{r}}}

\newcommand{\bx}{\boldsymbol{\textbf{x}}}
\newcommand{\bA}{\boldsymbol{\textbf{A}}}

\newcommand{\bH}{\boldsymbol{\textbf{H}}}
\newcommand{\bHbar}{\bar{\boldsymbol{\textbf{H}}}}

\newcommand{\bQ}{\boldsymbol{\textbf{Q}}}

\newcommand{\bM}{\boldsymbol{\textbf{M}}}

\newcommand{\bR}{\boldsymbol{\textbf{R}}}

\newcommand{\bT}{\boldsymbol{\textbf{T}}}

\newcommand{\bX}{\boldsymbol{\textbf{X}}}
\newcommand{\bY}{\boldsymbol{\textbf{Y}}}
\newcommand{\bphi}{\boldsymbol{\phi}}

\newcommand{\dr}{\,d\br}

\newcommand{\vKS}{v_{\rm{KS}}}
\newcommand{\vH}{v_{\rm{H}}}
\newcommand{\vN}{v_{\rm{N}}}
\newcommand{\vf}{v_{\rm{f}}}
\newcommand{\vext}{v_{\rm{ext}}}
\newcommand{\vxc}{v_{\rm{xc}}}
\newcommand{\HKS}{H_{\rm{KS}}}
\newcommand{\bAtilde}{\tilde{\textbf{A}}}

\newcommand{\modulus}[1]{\left\vert#1\right\vert}
\newcommand{\norm}[1]{\left\lVert#1\right\rVert}
\newcommand{\shp}[2]{N^{#1}_{#2}(\br)}
\usepackage{doi}
\usepackage{mathtools}
\usepackage{hyperref}
\hypersetup{
    colorlinks=true,
    linkcolor=magenta,
    filecolor=magenta,      
    urlcolor=cyan,
    }

\usepackage{subfiles}
\usepackage{graphicx}
\usepackage{graphics}
\usepackage{multirow}
\usepackage{amsxtra}
\usepackage{stmaryrd}
\usepackage{mathrsfs}
\usepackage{subcaption}
\usepackage{epsfig}
\usepackage{rotating}
\usepackage{setspace}
\usepackage{float}
\usepackage{amsfonts}
\usepackage{amsbsy}
\usepackage{amscd}
\usepackage{amsthm}
\usepackage{relsize}
\usepackage{color}
\usepackage{tabularx}
\usepackage{caption}
\usepackage{sidecap}
\usepackage{dcolumn}
\usepackage{footnote}
\usepackage{algorithm}
\usepackage{algorithmic}
\usepackage{breqn}

\definecolor{hellgruen}{rgb}{0.2,0.7,0.2}

\newcolumntype{M}[1]{>{\centering\arraybackslash}m{#1}}
\newcolumntype{N}{@{}m{0pt}@{}}
\title{Efficient all-electron time-dependent density functional theory calculations using an enriched finite element basis}

\author{Bikash Kanungo}
\affiliation{Department of Mechanical Engineering, University of Michigan, Ann Arbor, Michigan 48109, USA}
\author{Nelson D. Rufus}
\affiliation{Department of Mechanical Engineering, University of Michigan, Ann Arbor, Michigan 48109, USA}
\author{Vikram Gavini}
\affiliation{Department of Mechanical Engineering, University of Michigan, Ann Arbor, Michigan 48109, USA}
\alsoaffiliation{Department of Materials Science and Engineering, University of Michigan, Ann Arbor, Michigan 48109, USA}
\email{vikramg@umich.edu}

\begin{document}

\begin{abstract}
We present an efficient and systematically convergent approach to all-electron real-time time-dependent density functional theory (TDDFT) calculations using a mixed basis, termed as enriched finite element (EFE) basis. The EFE basis augments the classical finite element basis (CFE) with compactly supported numerical atom centered basis, obtained from atomic groundstate DFT calculations. Particularly, we orthogonalize the enrichment functions with respect to the classical finite element basis to ensure good conditioning of the resultant basis. We employ the second-order Magnus propagator in conjunction with an adaptive Krylov subspace method for efficient time evolution of the Kohn-Sham orbitals. We rely on \textit{a priori} error estimates to guide our choice of an adaptive finite element mesh as well as the time-step to be used in the TDDFT calculations. We observe close to optimal rates of convergence of the dipole moment with respect to spatial and temporal discretization. Notably, we attain a $50-100\times$ speedup for the EFE basis over the CFE basis. We also demonstrate the efficacy of the EFE basis for both linear and nonlinear response by studying the absorption spectrum in sodium clusters, the linear to nonlinear response transition in green fluorescence protein chromophore, and the higher harmonic generation in magnesium dimer. Lastly, we attain good parallel scalability of our numerical implementation of the EFE basis for up to $\sim1000$ processors, using a benchmark system of 50-atom sodium nanocluster.

\end{abstract}

\section{Introduction} \label{sec:intro}
An accurate description of electron excitations and dynamics under the influence of external stimuli is fundamental to our understanding of a range of physical and chemical processes. To that end, time-dependent density functional theory (TDDFT) offers a powerful tool by extending the key ideas of groundstate density functional theory (DFT) to electron excitations and dynamics. Analogous to the Hohenberg-Kohn theorem~\cite{Hohenberg1964} in groundstate density functional theory (DFT), TDDFT relies on the Runge-Gross ~\cite{Runge1984} and van Leeuwen~\cite{Leeuwen1999} theorems to establish, for an initial state, a one-to-one correspondence between the time-dependent external potential and the time-dependent electron density. This, in turn, provides a formally exact reduction of the complicated many-electron time-dependent Schr\"odinger equation to a set of effective single electron equations, called the time-dependent Kohn-Sham (TDKS) equations. At the heart of this simplification lies the exchange-correlation (XC) functional, which captures all the quantum many-electron interactions as a mean-field of the time-dependent electron density. Similar to DFT, in practice, TDDFT has remained far from exact due to the unavailability of the exact XC functional, thereby, necessitating the use of approximations. However, the available XC approximations has lent TDDFT a great balance of accuracy and efficiency, allowing the study of a wide array of time-dependent processes---optical~\cite{Appel2003} and higher-order responses~\cite{Gonze1989, vanGisbergen1997}, multi-photon ionization ~\cite{Tong1998, Chu2001, Tong2001, Telnov2009, Dauth2016}, electronic stopping~\cite{Hatcher2008, Yost2019, Pruneda2007, Correa2012, Schleife2015}, core electron excitations~\cite{Lopata2012, Attar2017, Pemmaraju2018, Pemmaraju2019, Yao2019}, surface plasmons~\cite{Ma2015, Peng2015}, higher-harmonic generation~\cite{Mrudul2020, Tancogne2017}, electron transport ~\cite{Stefanucci2004, Kurth2005}, charge-transfer excitations ~\cite{Jamorski2003, Stein2009}, dynamics of chemical bonds ~\cite{Burnus2005}, to name a few. 

While the XC approximation remains an unavoidable approximation in TDDFT, a typical TDDFT calculation also employs the pseudopotential approximation to attain greater computational efficiency by modeling the effect of the singular nuclear potential and the core electrons into a smooth effective potential, known as the pseudopotential. 
Despite tremendous success in predicting a wide range of materials properties, pseudopotentials remain sensitive to the choice of core-valence split and also tend to oversimplify the treatment of core electrons as chemically inert for various systems and conditions.
Within the context of groundstate DFT, pseudopotentials are known to have inaccurate predictions for the phase transition properties of transition metal oxides and semiconductors~\cite{Abu2000, Oganov2003, Koloren2007, Xiao2010} ; ionization potentials~\cite{Liu1998}, magnetizability~\cite{Schwerdtfeger2011}, and spectroscopic properties~\cite{Schwerdtfeger1995, Leininger1996, Schwerdtfeger2000} of heavy atoms; excited state properties~\cite{Gomez2008, Govoni2018} etc. 
More importantly, given that the construction of the pseudopotentials have happened in the context groundstate DFT, their deficiencies are expected to be more pronounced in TDDFT.
Several time-dependent processes involving the use of a strong external field rely on core electron excitations, wherein the use of pseudopotential approximation is impractical. 
Thus, all-electron TDDFT calculations are necessary for an accurate description of the time-dependent phenomena in such systems and conditions. Additionally, an efficient and robust all-electron TDDFT method can also aid in studying the transferability of various pseudopotentials for TDDFT calculations.     

The initial use of TDDFT relied on the linear response (LR) formulation, known as LR-TDDFT~\cite{Casida1995, Petersilka1996}, applicable to the perturbative regime (i.e., weak interaction between the external field and the electrons), wherein the first-order response functions (e.g., absorption spectrum) can be directly evaluated from the groundstate. 
Subsequently, the real-time formulation of TDDFT, known as RT-TDDFT, provided a generic framework to electronic dynamics in real-time, thereby, allowing to handle both perturbative and non-perturbative regimes (e.g., harmonic generation, electron transport) in a unified manner. 
This work pertains to the more general RT-TDDFT, and hence, we simply refer to RT-TDDFT as TDDFT. 
There exists a growing body of work on efficient numerical schemes for TDDFT calculations as extensions to widely used groundstate DFT packages, leveraging on the underlying spatial discretization. 
These include planewave basis in \texttt{QBox}~\cite{GPAW2005,Schleife2012}; atomic orbital basis in \texttt{Siesta}~\cite{Siesta2002,Takimoto2007}, \texttt{GPAW}~\cite{Kusima2015}, \texttt{NWChem}~\cite{NWChem2010,Lopata2011}, and \texttt{FHI-aims}~\cite{Blum2009,Hekele2021}; linearized augmented planewave (LAPW) basis  in \texttt{exciting}~\cite{Gulans2014,Pela2021} and \texttt{Elk}~\cite{elk}; and finite-difference (FD) based approaches in \texttt{Octopus}~\cite{Octopus2006} and \texttt{GPAW}~\cite{GPAW2005,Walter2008}. 
Among these available methods, planewave and FD based approaches are limited to pseudopoential calculations, owing to their lack of spatial adaptivity that is warranted to capture the sharp electronic fields in an all-electron calculation. The augmented planewave~\cite{Slater1964,Singh2006} family of methods, namely the augmented planewave (APW)~\cite{Loucks1967, Koelling1975}, linearized augmented planewave (LAPW)~\cite{Andersen1975, Wimmer1981, Weinert1982}, APW+lo (localized orbitals)~\cite{Sjostedt2000, Madsen2001, Gulans2014}, and LAPW+lo~\cite{Singh1991,Schwarz2002}, remedy the lack of adaptivity in planewaves by describing the electronic fields as products of radial functions and spherical harmonics inside muffin-tins (MTs) surrounding each
atom, and in terms of planewaves in the interstitial regions between atoms.  
Although efficient for all-electron calculations, the quality of the augmented planewave basis remains sensitive to various parameters, such as the choice of the MT radius, the core-valence split, the matching constraints at
MT boundary, the energy parameter used in constructing the radial functions, etc. 
Moreover, they inherit certain notable disadvantages of planewaves, such as their
restrictions to periodic boundary conditions and the limited parallel scalability owing to the the extended nature of planewaves. 
Above all, from a theoretical standpoint, the use of periodic boundary conditions limits the use of planewaves to only periodic external potentials, so as to satisfy the assumptions of Runge-Gross theorem. 
Alternatively, the periodic case needs to be handled
in a more generic and formal way through time-dependent current density functional
theory (TDCDFT)~\cite{Vignale2004}, where one uses the current density as the fundamental
quantity instead of the density.  
The atomic orbital basis provide an efficient description of the sharp electronic fields in all-electron DFT/TDDFT through atom specific analytical or numerical orbitals.
However, their lack of completeness limits systematic convergence, leading to significant basis set errors in groundstate properties, especially for metallic systems~\cite{Jensen2017a,Jensen2017b,Feller2018}.
More importantly, given that the atomic orbitals are constructed for groundstate DFT, the effects of incompleteness become more pronounced when they are employed in TDDFT calculations, leading to large basis set errors of $0.1-0.6$ eV in the excitation energies~\cite{Hekele2021}. 

The finite element (FE) basis~\cite{Brenner2007,Hughes2012}, comprising of local piecewise continuous polynomials, presents an alternative with several desirable features---completeness which guarantees systematic convergence, locality that affords good parallel scalability, ease of adaptive spatial resolution, and the ability to handle arbitrary boundary conditions. 
In the context of groundstate DFT, several past efforts~\cite{White1989, Tsuchida1998, Pask1999, Pask2001, Pask2005, Zhang2008, Suryanarayana2010, Fang2012, Bao2012, Motamarri2013, Motamarri2018, Das2019,Motamarri2020,Das2022} have established the promise of the FE basis. Particularly, recent efforts~\cite{Das2019,Motamarri2020,Das2022} at efficient and scalable FE based DFT calculations have outperformed planewaves by $5-10\times$, for pseudopotential based DFT calculations, and have been employed in various studies involving large-scale DFT calculations~\cite{motamarridna2020,ghosh2021,Lulu2022}. These developments have also enabled systematically converged inverse DFT calculations to obtain exchange-correlation potentials from electron densities~\cite{kanungo2019exact,kanungo2021}. 
In the context of TDDFT calculations, a few recent efforts~\cite{Lehtovaara2011,Bao2015, Kanungo2019} have established the competence of the FE basis. 
Notably, as shown in ~\cite{Kanungo2019}, for pseudopotential based TDDFT calculation, the FE basis significantly outperforms the widely used finite difference approach in \texttt{Octopus}.   
However, the success of the FE basis for pseudopotential calculations does not trivially extend to the all-electron case. 
As shown in~\cite{Motamarri2013, Kanungo2017}, for all-electron DFT calculations, the FE basis remains an order of magnitude or more inefficient than the gaussian basis in computational time, owing to the requirement of large number of basis functions to capture the sharp variations in electronic fields near the nuclei in an all-electron calculation. 
As will be shown in the work, this shortcoming of FE basis for the all-electron case, as expected, also extends to TDDFT calculations. 

Given the various shortcomings of existing basis sets for all-electron TDDFT calculations, an efficient, systematically convergent, and highly parallelizable basis is desirable. 
The current work seeks to address this gap by extending the ideas of enriched finite element (EFE) basis, recently developed in the context of DFT calculations~\cite{Kanungo2017, Rufus2021, Rufus2022}, to TDDFT. 
The EFE basis is a mixed basis comprising of the classical/standard FE (CFE) basis along with compactly supported numerical atom-centered basis, known as enrichment functions. 
In effect, the EFE basis combines the efficiency of the atomic orbitals
to capture the essential features of the electronic fields near the nuclei with the completeness of the FE basis. 
As demonstrated in~\cite{Kanungo2017, Rufus2021}, the EFE basis significantly outperforms the gaussian basis for nanoclusters as well as competes with the LAPW basis for solids. 
In this work, we present an efficient EFE basis based formulation of TDDFT and demonstrate its resulting advantages over the CFE basis. 
In particular, we employ an \textit{apriori} mesh adaption strategy to obtain an efficient CFE basis~\cite{Kanungo2019}. 
Additionally, we leverage on a full-discrete error analysis of the problem~\cite{Kanungo2019}, in
the context of second-order Magnus propagator, to obtain a economic choice for the time-step used in the propagation. 
Lastly, we apply an adaptive Krylov subspace method to efficiently compute the action of the Magnus propagator, given as an exponential operator, on the Kohn-Sham orbitals. 

We study various numerical aspects of the EFE basis for all-electron TDDFT calculations, for both weak and strong perturbations. 
We, first, demonstrate close to optimal rates of convergence of the dipole moment with respect to both spatial and temporal discretization, using carbon monoxide (CO) under weak perturbation as a benchmark system. 
Subsequently, we examine the accuracy and performance of the EFE basis for larger all-electron TDDFT calculations using methane and benzene molecules as well as sodium nanoclusters of increasing sizes as our benchmark systems. We attain a remarkable $50-100\times$ speedup over the CFE basis, while being commensurate with an accuracy of $<1$ mHa in the excitation energy. 
We investigate the competence of the EFE basis for strong perturbation through a comparative study of the dipole response and absorpotion spectrum of the green fluorescent protein (GFP) chromophore subjected to electric field of varying strength.
We further establish the efficacy of the EFE basis for strong perturbation by studying the higher harmonic generation in magnesium dimer.
Lastly, we demonstrate good parallel scalability of our implementation up to $\sim1000$ processors, for a benchmark 50-atom sodium nanocluster discretized using $\sim1.7$ million basis functions. 

The rest of the paper is organized as follows: Sec.~\ref{sec:methods} describes the theory and the methodology, Sec.~\ref{sec:results} presents the numerical results, and, finally, we summarize the findings and outline the future scope of the work in Sec.~\ref{sec:conclusion}.

\section{Theory and Methodology} \label{sec:methods}
\subsection{Time dependent Kohn-Sham equations} \label{sec:TDKS}
TDDFT reduces the many-electron time dependent Schr\"odinger equation to an auxiliary system of non-interacting electrons which yields the same time dependent electron density, $\rho(\br,t)$, as the interacting system and whose evolution is prescribed by a set of effective single electron equations, called the time dependent Kohn-Sham (TDKS) equations. 
The TDKS equations, in atomic units, are given as
\begin{equation} \label{eq:TDKS}
\begin{split}
    i\frac{\partial{\psi\al(\br,t)}}{\partial{t}} &= \HKS[\rho](\br,t;\bR)\psi\al(\br,t)
    \coloneqq \left[-\frac{1}{2}\nabla^2+ \vKS[\rho](\br,t;\bR)\right]\psi\al(\br,t) \\
    & \coloneqq \left[-\frac{1}{2}\nabla^2+ \vext(\br,t;\bR) + \vH[\rho](\br,t) + \vxc[\rho](\br,t)\right]\psi\al(\br,t)\,. 
\end{split}
\end{equation}
In the above, $\HKS[\rho](\br,t;\bR)$ and $\psi\al(\br,t)$ represent the time-dependent Kohn-Sham Hamiltonian and orbitals, respectively; $\alpha$ is the index for the $N_e$ electrons in the system; $\bR=\{\bR_1,\bR_2,...,\bR_{N_a}\}$ is the collective representation for the positions of the $N_a$ atoms in the system. 
The electron density, $\rho(\br,t)$, is defined by the the Kohn-Sham orbitals as
\begin{equation}
    \rho(\br,t)=\sum_{\alpha=1}^{N_e}|\psi\al(\br,t)|^2\,.
\end{equation}
For simplicity, we consider only spin-unpolarized systems. However, the ideas presented can be easily extended to spin-polarized systems. This work does not account for relativistic effects, which will be the subject of a future work.    
The external potential, $\vext(\br,t;\bR)$, comprises of the nuclear potential ($\vN(\br;\bR$)) and the potential corresponding to an external field ($\vf(\br,t)$). 
The nuclear potential is given as
\begin{equation} \label{eq:VN}
    \vN(\br;\bR) = -\sum\limits_{I=1}^{N_a}\frac{Z_I}{|\br-\bR_I|}\,,
\end{equation}
where $Z_I$ is the atomic number of the $I-$th atom. $\vf(\br,t)$ is usually defined as a monochromatic laser pulse of the form 
\begin{equation} \label{eq:VField}
    \vf(\br,t)=-\mathbf{E_0}(t)\cdot\br\,,
\end{equation}
where $\mathbf{E_0}(t)$ represents the time-dependent electric field. 
The Hartree potential, $\vH[\rho](\br,t)$, denotes the classical electrostatic potential corresponding to the electron density, given as
\begin{equation} \label{eq:VH}
    \vH[\rho](\br,t)=\int{\frac{\rho(\br',t)}{|\br-\br'|}\dr'}\,.
\end{equation}
The exchange-correlation potential, $\vxc[\rho](\br,t)$, is the mean-field potential that accounts for the quantum many-electron interactions. 
Although, in general, the exchange-correlation potential is known to be nonlocal in both space and time~\cite{Vignale1995, Maitra2002,Marques2006} along with initial state dependence, most widely used approximations use locality in time (adiabatic approximation) and non-dependence on the initial many-electron wavefunction, for lack of knowledge of the time nonlocality.
As a result, it is common to employ the existing exchange-correlation approximations used in ground-state DFT. 
In this work, we use the adiabatic local-density approximation (ALDA)~\cite{Gross1985}, which is local in both space and time. 
Specifically, we use the Ceperley-Alder form \cite{Ceperley1980}.

We note that both $\vN$ and $\vH$ are extended in real space, and for an evaluation in real-space formulations, they can be recast as solutions to the following Poisson equations~\cite{Pask1999, Pask2005, Suryanarayana2010, Motamarri2013, Motamarri2012}:
\begin{subequations} \label{eq:Poisson}
  \begin{equation} \label{eq:Hartree_possion}
    -\frac{1}{4\pi}\nabla^2\vH(\br,t)=\rho(\br,t)\,,\quad \vH(\br,t)\rvert_{\partial\Omega}=f(\br,\bR) \,,
  \end{equation}    
  \begin{equation} \label{eq:Nuclear_poisson}
      -\frac{1}{4\pi}\nabla^2\vN(\br;\bR) = b(\br,\bR) \quad  \vN(\br)\rvert_{\partial\Omega}=-f(\br,\bR)\,,
  \end{equation}
\end{subequations}
where $b(\br,\bR) = -\sum\limits_{I=1}^{N_a}{Z_I\delta(\modulus{\br-\bR_I})}$ with $\delta(x)$ denoting a Dirac delta distribution; $f(\br,R)=\sum_{I=1}^{N_a}\frac{Z_I}{\modulus{\br-\bR_I}}$; $\partial \Omega$ denotes the boundary of a sufficiently large bounded domain $\Omega \in \mathcal{R}^3$.

\subsection{Enriched finite element (EFE) discretization} \label{sec:EFE}
In this section, we present the EFE discretization of the TDKS equations. 
At the heart of the EFE discretization lies the augmentation of the CFE basis with compactly supported numerical atomic functions, termed as enrichment functions. 
As a result, we account for the sharp variations in the electronic fields close to nuclei, largely,
through the enrichment functions and mitigate the need for a refined classical finite element mesh. 
The initial attempt at an EFE basis~\cite{Kanungo2017} for groundstate DFT used smoothly truncated single atom orbitals and potentials as the enrichment functions, which resulted in a staggering  $50-100\times$ speedup over the CFE basis. 
However, such enrichment functions are susceptible to being linearly dependent on the CFE basis, particularly while using refined FE meshes, affecting the accuracy and robustness of the resulting EFE basis. 
To that end, we developed an orthogonalized EFE basis~\cite{Rufus2021}, wherein the enrichment functions are orthogonalized with respect to the underlying CFE basis, without comprising on the locality of the resultant basis. 
This work employs the more robust orthogonalized EFE basis.
Hereafter, for succinctness, we omit the qualifier orthogonalized and refer to the orthogonalized EFE basis simply as EFE basis. 
The EFE discretization of the Kohn-Sham orbital, $\psi\al(\br,t)$, is given as
\begin{equation} \label{eq:PsiEFE}
    \psi\al^h(\br,t)=\underbrace{\sum_{i=1}^{n_h}\shp{C}{i}\psi_{\alpha,i}^C(t)}_{\text{Classical}} + \underbrace{\sum_{I=1}^{N_a}\sum_{j=1}^{n_I}\shp{E,\psi}{j,I}\psi_{\alpha,j,I}^{E}(t)}_{\text{Enriched}}\,,
\end{equation}
where the superscript `$h$' denotes a discrete field; \{$\shp{C}{i}$\} and $\{\psi_{\alpha,i}^C(t)\}$ denote the classical FE basis functions and their corresponding time-dependent coefficients; similarly, \{$\shp{E,\psi}{j,I}$\} and $\{\psi_{\alpha,j,I}^E(t)\}$ denote the enrichment basis functions and their respective time-dependent coefficients.
In the above form, the $I$\textsuperscript{th} atom at $\bR_I$, contributes $n_I$ enrichment functions, each centered around $\bR_I$. 
The enrichment function \{$\shp{E,\psi}{j,I}$\} is expressed as
\begin{equation} \label{eq:EFESplit}
    \shp{E,\psi}{j,I} = \shp{A,\psi}{j,I} - \shp{B,\psi}{j,I}\,,
\end{equation}
where $\shp{A,\psi}{j,I}$ and $\shp{B,\psi}{j,I}$ are the atomic and orthogonalizing parts, respectively. The atomic part, $\shp{A,\psi}{j,I}$, is given as
\begin{equation} \label{eq:EFEAtomic}
    \shp{A,\psi}{j,I} = \psi_{nlm,I}(|\bx-\bR_I|,\beta_{\bR_I},\gamma_{\bR_I})u(|\bx-\bR_I|,r_0,s)\,,
\end{equation}
where $\psi_{nlm,I}$ is an atomic groundstate Kohn-Sham orbital indexed by the principal quantum number $n$, azimuthal quantum number $l$, and magnetic quantum number $m$, for an isolated atom of the atom type of the $I^{\text{th}}$ atom, defined in spherical coordinates. The function $u(r,r_0,s)$ is a smooth cutoff function, parameterized by a cutoff radius $r_0$ and smoothness factor $s$. 
Typically, we include all the occupied and a few unoccupied groundstate single atom orbitals as the enrichment functions. The orthogonalizing part, $\shp{B,\psi}{j,I}$, is given as a linear combination of the underlying CFE basis, i.e.,
\begin{equation} \label{eq:EFEOrtho}
    \shp{B,\psi}{j,I} = \sum_{l=1}^{n_h} c_{j,I,l}\shp{C}{l}\,,
\end{equation}
where $c_{j,I,l}$ are evaluated such that they satisfy the orthogonality condition $\int_{\Omega}\shp{E,\psi}{j,I}\shp{C}{i}\dr=0$.  
We refer to~\cite{Rufus2021} for a detailed description on $\shp{A,\psi}{j,I}$, $\shp{B,\psi}{j,I}$, and $h(r,r_0,s)$. Hereafter, for compact notation, we combine the $\{j,I\}$ indices in $\shp{E,\psi}{j,I}$ and $\psi_{\alpha,j,I}$ into a single index $\nu$.  

Using the EFE discretization of Eq.~\ref{eq:PsiEFE} in the TDKS equations (cf. Eq.~\ref{eq:TDKS}) yields
\begin{equation} \label{eq:TDKSDiscrete}
    i\bM \dot{\bpsi}\al(t)= \bH \bpsi\al(t)\,.
\end{equation}
where $\bH$ and $\bM$ are the discrete Kohn-Sham Hamiltonian matrix and the overlap matrix, respectively, and $\bpsi\al$ denotes the vector containing the $\psi_{\alpha,i}^C(t)$ and $\psi_{\alpha,\nu}^E(t)$ coefficients of Eq.~\ref{eq:PsiEFE}. The Kohn-Sham Hamiltonian $\bH$ is given by
\begin{equation} \label{eq:Hmn}
H_{mn} = \frac{1}{2}\int_{\Omega}\nabla N_m(\br)\cdot \nabla N_n(\br) \dr + \int_{\Omega}\vKS[\rho](\br,t;\bR)N_m(\br)N_n(\br)\dr\,, 
\end{equation}
where $N_m(\br)$ and $N_n(\br)$ are generic representations for $\shp{C}{i}$ and $\shp{E,\psi}{\nu}$. The overlap matrix $\bM$, owing to the orthogonality between $\shp{C}{i}$ and $\shp{E,\psi}{\nu}$, has the following block-diagonal structure
\begin{equation} \label{eq:M}
\bM=\left[
\begin{array}{c|c}
\bM^{\textbf{cc}}  & 0 \\ \hline
 0 &  \bM^{\textbf{ee}}
\end{array}\right]\,,
\end{equation}
where $\bM^{\textbf{cc}}_{jk}=\int_{\Omega}\shp{C}{j}\shp{C}{k}\dr$ denotes the overlap between two CFE basis functions and $\bM^{\textbf{ee}}_{\mu,\nu}=\int_{\Omega}\shp{E,\psi}{\mu}\shp{E,\psi}{\nu}$ denotes the overlap between two enrichment functions. 
Noting the fact that $\bM$ is a positive definite matrix with a unique positive definite square root, $\bM^{1/2}$, we reformulate Eq.~\ref{eq:TDKSDiscrete} as
\begin{equation} \label{eq:TDKSDiscrete2}
    i\dot{\bpsibar}\al(t)= \bHbar \bpsibar\al(t)\,,
\end{equation}
where $\bpsibar\al(t) = \bM^{1/2}\bpsi\al(t)$ and $\bHbar = \bM^{-1/2}\bH\bM^{-1/2}$ are the representation of $\psi\al(\br,t)$ and $\HKS$ in a L\"owdin orthonormalized basis ~\cite{Lowdin1950}.
We emphasize that the above transformation of the discrete TDKS equation is crucial to our use of the Magnus propagator (see Sec.~\ref{sec:magnus}).

Similar to $\psi\al(\br,t)$, the EFE discretization of the electrostatic potential (both nuclear and Hartree potentials), $\phi(\br,t)$, can be written as
\begin{equation} \label{eq:PhiEFE}
    \phi^h(\br,t)=\underbrace{\sum_{j=1}^{n_h}\shp{C}{j}\phi_{j}^C(t)}_{\text{Classical}} + \underbrace{\sum_{I=1}^{N_a}\shp{E,\phi}{I}\phi_{I}^E(t)}_{\text{Enriched}}\,.
\end{equation}
We note that for the nuclear potential, $\phi(\br,t)$ does not have any time dependence. 
In the above, the enrichment function $\shp{E,\phi}{I}$, analogous to $\shp{E,\psi}{\nu}$, comprises of an atomic part ($\shp{A,\phi}{I}$) and an orthogonalizing part ($\shp{B,\phi}{I}$). 
The atomic part, $\shp{A,\phi}{I}$, is given as a smoothly truncated spherically symmetric electrostatic potential (nuclear or Hartree) of an isolated atom of the same type as located at $\bR_I$. 
On the other hand, the orthogonalizing part, $\shp{B,\phi}{I}$, is given as a linear combination of the CFE basis \{$\shp{C}{i}$\}, such that it guarantees the orthogonality condition: $\int_\Omega\shp{E,\phi}{I}\shp{C}{i}\dr=0$.  Using Eq.~\ref{eq:PhiEFE} in Eq.~\ref{eq:Poisson} yields
\begin{equation} \label{eq:DiscretePoisson}
    \bA \bphi(t) = \bd(t)\,,
\end{equation}
where $\bA_{mn}=\frac{1}{4\pi}\int_{\Omega} \nabla N_m(\br)\cdot N_n(\br)$, with  $N_m(\br)$ and  $N_n(\br)$ being generic representation of $\shp{C}{i}$ and $\shp{E,\phi}{I}$, is the discrete Laplace operator; and $\bd_m(t)=\int_{\Omega}\rho(\br,t)N_m(\br)$ for the Hartree potential or $\bd_m(t)=\int_{\Omega}b(\br;\bR)N_m(\br)$ for the nuclear potential.

\subsection{Spectral finite elements}
We remark that the L\"owdin transformation in Eq.~\ref{eq:TDKSDiscrete2} warrants an efficient means of evaluating $\bM^{1/2}$ and $\bM^{-1/2}$. 
To that end, we use spectral finite elements in conjunction with Gauss-Lobatto-Legendre (GLL) quadrature rule, the combination of which renders classical-classical block ($\bM^{\textbf{cc}}$) block of $\bM$ diagonal, and, hence, allows for trivial evaluation of $\left(\bM^{\textbf{cc}}\right)^{1/2}$ and $\left(\bM^{\textbf{cc}}\right)^{-1/2}$. 
We refer to~\cite{Motamarri2013} for an elaborate discussion on spectral finite elements. 
The enriched-enriched block ($\bM^{\textbf{ee}}$), on the other hand, is a small diagonally dominant matrix of size $\sim N_e\times N_e$, and, hence, $\left(\bM^{\textbf{ee}}\right)^{1/2}$ and $\left(\bM^{\textbf{ee}}\right)^{-1/2}$ can be easily evaluated through either eigenvalue decomposition or Newton-Schultz methods ~\cite{Jansik2007,Nikalsson2004,Higham1997}. 
Subsequently, $\bM^{1/2}$ and $\bM^{-1/2}$ can be written as
\begin{equation} \label{eq:MSquareRoot}
\bM^{1/2}=\left[
\begin{array}{c|c}
(\bM^{\textbf{cc}})^{1/2} & 0 \\ \hline
0  &   \left( \bM^{\textbf{ee}} \right)^{1/2}
\end{array}\right]\,, \quad
\bM^{-1/2}=\left[
\begin{array}{c|c}
(\bM^{\textbf{cc}})^{-1/2} & 0 \\ \hline
0  &   \left( \bM^{\textbf{ee}} \right)^{-1/2}
\end{array}\right]\ \,.
\end{equation}

\subsection{Adaptive quadrature}\label{sec:adaptiveQuadrature}
The enrichment functions, $\shp{E,\psi}{\nu}$ and $\shp{E,\phi}{I}$, being characterized by sharp gradients or oscillations near the atoms, warrant the use of a high density of quadrature points near the atoms for an accurate evaluation of the integrals involving them. 
On the other hand, given that the enrichment functions have a small compact support, a lower quadrature density may suffice in regions farther away from atoms. 
We strike a good balance of accuracy and efficiency by using an adaptive quadrature, wherein we recursively refine each finite element until a set of trial integrals, involving the enrichment functions, attain convergence. To this end, we adopt the adaptive quadrature scheme proposed in ~\cite{Mousavi2012,Kanungo2017} and refer to these works for the details.

\subsection{Magnus Propagator} \label{sec:magnus}
We now discuss the time discretization of Eq.~\ref{eq:TDKSDiscrete2} in the context of second-order Magnus propagator. In the Magnus \textit{ansatz}, the solution to Eq. ~\ref{eq:TDKSDiscrete2} is given as
\begin{equation} \label{eq:MagnusAnsatz} 
    \bpsibar\al(t) = \text{exp}(\bA(t)) \bpsibar\al(0)\,, \qquad \forall t \geq 0 \,,
\end{equation}
where $\text{exp}(\bA(t))$ is called the Magnus propagator with $\bA(t)$ given explicitly as ~\cite{Blanes2009, Hochbruck2003}
\begin{equation} \label{eq:MagnusExpansion}
    	\bA(t) = \int_0^t -i\bHbar(\tau)d\tau - \frac{1}{2}\int_0^t\left[\int_0^{\tau} -i\bHbar(\sigma)d\sigma,-i\bHbar(\tau)\right]d\tau 
    	+ \ldots\,,
\end{equation}
with $[\bX,\bY]=\bX\bY-\bY\bX$ being the commutator. Given the difficulty in accurately resolving the implicit dependence of $\bHbar(t)$ on $\rho(\br,t)$, for practical use, we rewrite the Magnus propagator as 
\begin{equation} \label{eq:MagnusAnsatzSplit}
    \text{exp}(\bA(t)) = \prod_{n=1}^{N}\exp(\bA_n) \,,
\end{equation}
where $\bA_n$ is given by Eq.~\ref{eq:MagnusExpansion} with the limits of integration modified to $[t_{n-1},t_{n}]$.
In practice, we use an approximation to the exact $\bA_n$, denoted as $\bAtilde_n$, which involves a truncation of the Magnus expansion (Eq.~\ref{eq:MagnusExpansion}) and an approximation for the time integrals in the truncated Magnus expansion. 
Using the first $p$ terms in the Magnus expansion results in a time-integration scheme of order $2p$. In this work, we use the second-order Magnus propagator, i.e.,
obtained by truncating the Magnus expansion after the first term. Additionally, we use a mid-point integration rule to evaluate $\int_{t_{n-1}}^{t_n}-i\bHbar(\tau)d\tau$. As a result, given a set of Kohn-Sham orbitals $\{\bpsibar_1(t),\bpsibar_2(t),\ldots,\bpsibar_{N_e}(t)\}$ defining the density $\rho(\br,t)$, we can write
\begin{equation} \label{eq:ApproxMagnus}
        \bpsibar\al(\br,t_{n-1}+\Delta t) \approx e^{\bAtilde_n}\bpsibar\al(t_{n-1}) = e^{-i \bHbar\left[\rho\left(t_{n-1}+\frac{\Delta t}{2}\right)\right]\Delta t}\bpsibar\al(t_{n-1})\,,
\end{equation}
where $\Delta t = t_n-t_{n-1}$ and $\bHbar\left[\rho\left(t_{n-1}+\frac{\Delta t}{2}\right)\right]$ is the Kohn-Sham Hamiltonian described at the future time instance $t_{n-1}+\Delta t/2$. 
$\bHbar\left[\rho\left(t_{n-1}+\frac{\Delta t}{2}\right)\right]$, being dependent on a future instance of the density, is evaluated either by an extrapolation of $\bHbar$ using $m (> 2)$ previous steps or by a second (or higher) order predictor-corrector scheme.
In this work, we use the second-order predictor-corrector scheme presented in~\cite{Cheng2006} (also see~\cite{Kanungo2019} for details).

\subsection{Adaptive Krylov subspace}
The use of the second-order Magnus propagator, as defined in Eq.~\ref{eq:ApproxMagnus}, requires an efficient means of evaluating the action of an exponential operator ($\text{exp}({{\bAtilde}_n})$)  on a vector ($\bpsibar\al(t_{n-1})$). While direct evaluation of $\text{exp}({{\bAtilde}_n})$ is computationally prohibitive beyond small sizes, one can employ subspace projection methods to approximate the action of the exponential operator on a vector. To that end, we use an adaptive Krylov subspace method, wherein we approximate $\text{exp}({{\bAtilde}_n})\bpsibar$ as
\begin{equation} \label{eq:AdaptiveKrylov}
 \text{exp}({{\bAtilde}_n})\bpsibar \approx \norm{\bpsibar}\bQ_k \text{exp}({\bQ_k^\dagger \bAtilde_n \bQ_k}) \hat{e}_1 = \norm{\bpsibar}\bQ_k \text{exp}({\bT_k}) \hat{e}_1\,,     
\end{equation}
where $\bQ_k=\{\bq_1,\bq_2,\ldots,\bq_k\}$ denotes $k$ orthonormal set of vectors (known as Lanczos vectors) that span the Krylov subspace $\mathcal{K}_k\{\bAtilde_n,\bpsibar\}=\{\bpsibar,\bAtilde_n\bpsibar,\bAtilde_n^2\bpsibar,\ldots,\bAtilde^{k-1}\bpsibar\}$; $\bT_k=\bQ_k^\dagger \bAtilde_n \bQ$ is a tridiagonal matrix; and $\hat{e}_1$ is the first unit vector in $\mathbb{C}^k$. 
Thus, the problem is now reduced to the evaluation of $\text{exp}(\bT_k)$, where $\bT_k$ is a small matrix  $k \times k$ matrix, and, hence, $\text{exp}(\bT_k)$ can be inexpensively computed either through Taylor expansion or exact eigenvalue decomposition of $\bT_k$. 
A distinct advantage of the Krylov subspace approach is that the error, $\epsilon_k$, incurred in the above approximation can be estimated as ~\cite{Hochbruck1998}
\begin{equation}\label{eq:AdaptiveKrylovErr}
\epsilon_k=\norm{\text{exp}({\bAtilde_n})\bpsibar - \norm{\bpsibar}\bQ_k \text{exp}({\bT_k})e_1} \approx \beta_{k+1,k}\norm{\bpsibar}\modulus{\left[\text{exp}({\bT_k})\right]_{k,1}}\,,
\end{equation} 
where $\beta_{k+1,k}$ is the $(k+1,k)$ entry of $\bT_{k+1}=\bQ_{k+1}^\dagger\bAtilde_n\bQ_{k+1}$. 
As a result, the above relation offers a systematically convergent, efficient and adaptive recipe to evaluate the action of the second-order Magnus propagator on the Kohn-Sham orbitals.

\subsection{Efficient mesh adaption and temporal discretization} \label{sec:efficient}
A salient feature of the FE basis is its adaptive spatial resolution, thereby, allowing for finer resolution near the atoms and coarser resolution away from the atoms. 
The benefits of adaptive resolution is even more pronounced in the case of TDDFT, which warrants the use of large simulation domains to mitigate any spurious artifacts from reflections of the boundary of the simulation domain. 
However, an efficient and reliable choice of an adaptive FE mesh is useful to be guided by error analysis. 
To that end, we employ the \textit{a priori} mesh adaption techniques presented in~\cite{Kanungo2019} to generate the underlying CFE basis of our EFE basis. 
The key idea is to minimize the semi-discrete (discrete in space, but continuous in time) error in the dipole moment of the system with respect to the mesh-size distribution ($h(\br)$).   
Briefly, the mesh adaption strategy is as follows: (i) for a fixed number of finite elements ($N_{\text{elem}}$), we use the ground-state atomic orbitals and electrostatic potentials to determine the optimal mesh size distribution expression presented in ~\cite{Kanungo2019} (see Eq. 34 in the reference); (ii) we increase the value of $N_{\text{elem}}$ until chemical accuracy in the quantities of interest (e.g., groundstate energy, excitation energy) is attained.
Typically, we conduct the mesh adaption exercise for small representative systems (atoms or small molecules) to determine the characteristic $h(\br)$ that attains chemical accuracy. Subsequently, for larger systems, we use the $h(\br)$ from the nearest atom.

Similar to an efficient mesh adaption, we turn to the full-discrete (discrete in both space and time) error analysis presented in~\cite{Kanungo2019} for an economic choice of the time-step ($\Delta t$).
The key aspect of the error analysis is to relate, for a given level of accuracy, the optimal $\Delta t$ to the spatial discretization error. 
In practice, we first obtain the optimal $\Delta t$ from atomic TDDFT calculations, as per the full-discrete error estimate in~\cite{Kanungo2019} (see Eq. 35 in the reference).
Subsequently, for a multi-atom case we use the least $\Delta t$ obtained for each of its constituent atomic species.

\section{Results and Discussion} \label{sec:results}
We discuss various numerical aspects of the EFE basis for all-electron TDDFT, ranging from accuracy and rate of convergence to computational efficiency and parallel scalability. 
All the calculations use the groundstate Kohn-Sham orbitals as the initial states.
All our calculations are conducted on a parallel computing cluster with the following configuration: Intel Xeon Gold 6154 (Skylake) CPU nodes with 36 processors (cores) per node, 180 GB memory per node, and InfiniBand HDR100 networking networking between all nodes for fast MPI communications. The enrichment functions contributed by each atom in a system comprise of all its occupied Kohn-Sham orbitals as well as its Kohn-Sham orbitals belonging to the lowest unoccupied shell, obtained from groundstate DFT calculation. For example, for O, we include its occupied orbitals of $1s$, $2s$, and $2p$ as well as all the orbitals from its lowest unoccupied shell (i.e., $3s$, $3p$, and $3d$). 

\subsection{Rate of convergence} \label{sec:convergence}
\begin{figure}[h!]
\begin{center}
\includegraphics[scale=1.2]{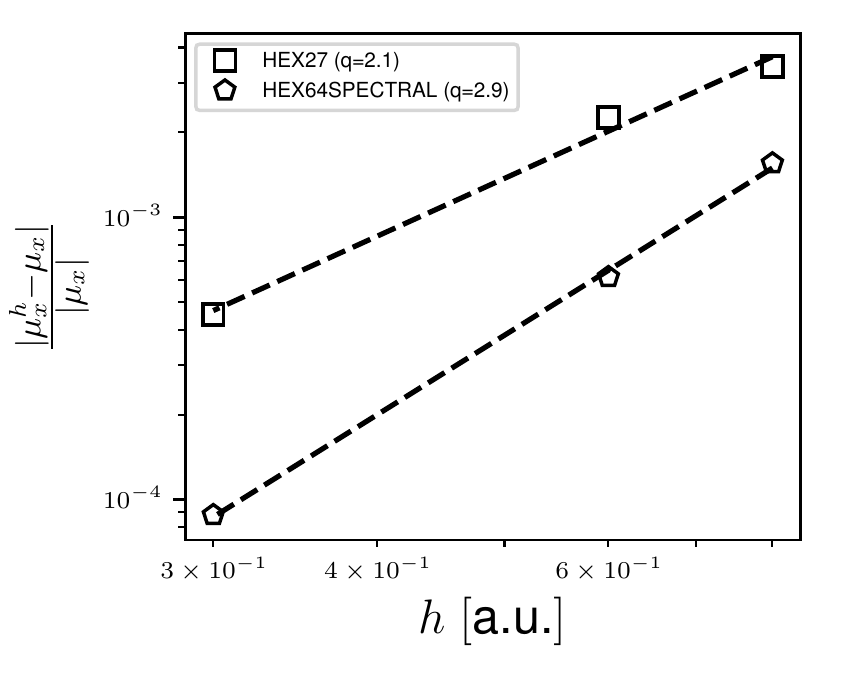}
\caption{\small Rates of convergence with respect to mesh size for CO.}
\label{fig:COSpatial}
\end{center}
\end{figure}

We study the rates of convergence of the dipole moment afforded by the EFE basis with respect to both spatial (FE mesh size $h$) and temporal discretization (time-step $\Delta t$), using a CO molecule of bond length 2.4 a.u.\ as a benchmark system. 
In order to reliably study the rates of convergence with respect to the mesh size, we need to mimic a semi-discrete (discrete in space but continuous in time) error analysis, so as to make the spatial discretization error to be the dominant source of error. 
To that end, we use a small time-step of $\Delta t =10^{-3}$ and small tolerance of $10^{-12}$ for the Krylov subspace projection error (Eq.~\ref{eq:AdaptiveKrylovErr}). 
A large cubical domain of side 50 a.u.\ is chosen to ensure that the electron density decays to zero on the domain boundary,
allowing us to impose Dirichlet boundary condition on the time-dependent Kohn-Sham orbitals and the Hartree potential. 
We consider two different orders ($p$) of finite elements---HEX27 ($p=2$) and HEX64SPECTRAL ($p=3$). 
For each $p$, we consider a sequence of increasingly refined meshes using the recipe presented in ~\ref{sec:efficient} with increasing values of $N_{\text{elem}}$.  
For all the meshes, we first, obtain the ground-state and then excite the system using a Gaussian electric field of the form $\textbf{E}_0(t)=\kappa e^{(t-t_0)^2/2s^2} \hat{\bx}_1$, with $\kappa=2\times 10^{-5}$ a.u., $t_0=3.0$ a.u., $s=0.2$ a.u.\, and $\hat{\bx}_1$ denoting the unit vector along $x$-direction. We note that the semi-discrete error in the dipole moment, as a function of the mesh size ($h$), can be expressed as~\cite{Kanungo2019}
\begin{equation}\label{eq:DipoleSemi}
  \frac{\modulus{\mu_{x}(t)-\mu_x^h(t)}}{\modulus{\mu_{x}(t)}}=Ch^q\,,
\end{equation}
where $\mu_x(t)$ and $\mu_x^h(t)$ are the continuum and discrete values of the $x$-component of the dipole moment at time $t$, $C$ is a mesh-independent constant, and $q$ is the rate of convergence. 
We first evaluate $\mu_x(t)$ using the EFE basis with a highly refined HEX125SPECTRAL ($p=4$) finite-element mesh. 
Subsequently, $C$ and $q$ are obtained by fitting the above relation to a set of $\mu_x^h(t)$ and $h$.
Fig.~\ref{fig:COSpatial} presents the relative semi-discrete error in the dipole moment at $t=5$ a.u.\, (1 a.u.\ = 0.024188 fs). 
As evident, the numerical rates of convergence are in close agreement with the theoretical rate of $\mathcal{O}(h^{p})$, where $p$ is the order of the FE basis ($p=2$ for HEX27 and $p=3$ for HEX64SPECTRAL).  

\begin{figure}[h!]
\begin{center}
\includegraphics[scale=1.2]{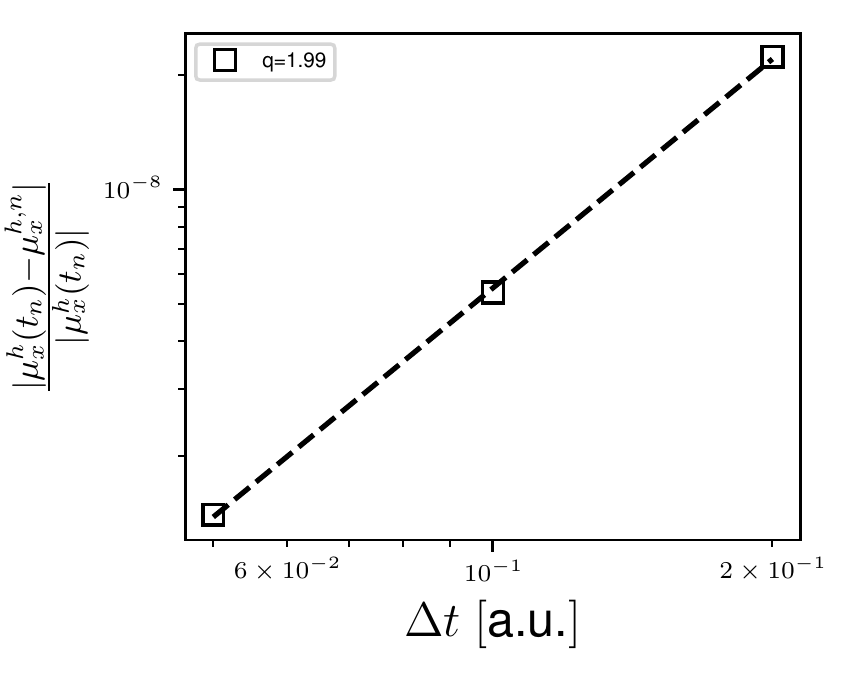}
\caption{\small Rates of convergence with respect to $\Delta t$ for CO.}
\label{fig:COTemporal}
\end{center}
\end{figure}

We, next, turn to the study the rate of convergence of the dipole moment with respect to temporal discretization. 
To this end, we used a finite element mesh with a sufficiently refined HEX125SPECTRAL mesh, such that it affords $10^{-4}$ relative error with respect to spatial discretization. 
We use the same Gaussian electric field as above and propagate the ground-state Kohn-Sham orbitals using second-order Magnus propagator with different $\Delta t$. 
The full-discrete (discrete both in space and time) error in the dipole moment at time $t_n$ is given as~\cite{Kanungo2019}
\begin{equation}\label{eq:DipoleFull}
    \frac{\modulus{\mu^{h}_x(t_n)-\mu^{h,n}_{x}}}{\modulus{\mu^{h}_x(t_n)}} = C (\Delta t)^q \,,
\end{equation}
where $\mu^{h}_x(t_n)$ and $\mu^{h,n}_{x}$ are the semi-discrete and full-discrete values of the $x$-component of the dipole moment at time $t_n$, $C$ is a constant, and $q$ is the rate of convergence.   
Fig. ~\ref{fig:COTemporal} depicts the rate of convergence of the dipole moment with respect to $\Delta t$ at $t_n=5.0$ a.u.\,. We attain a numerical rate of convergence of $q=1.99$, which is in good agreement with the quadratic rate of convergence for second-order Magnus propagator. 

\subsection{Weak perturbation on methane and benzene}\label{sec:methane_benzene}
\begin{figure}[h!]
\begin{center}
\includegraphics[scale=1.2]{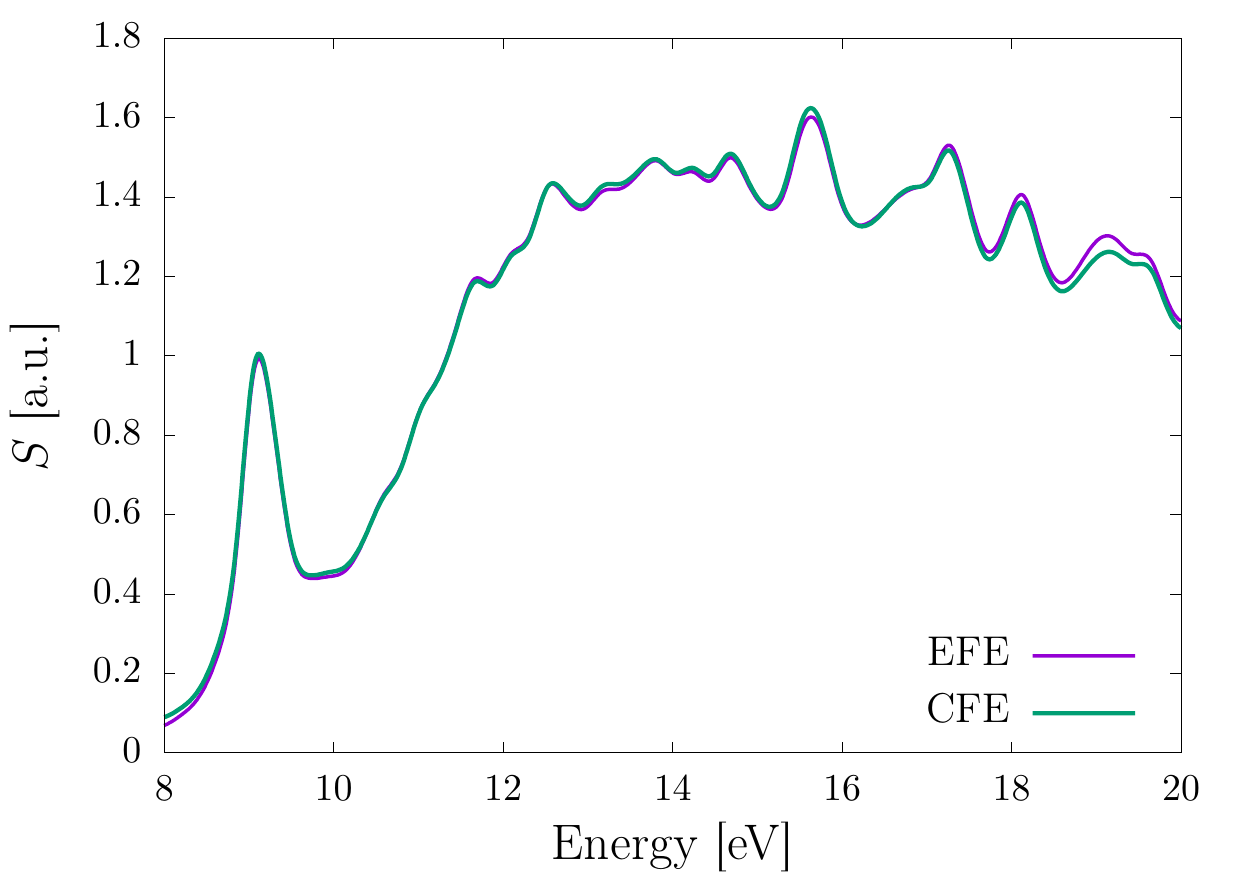}
\caption{\small Absorption spectrum of methane.}
\label{fig:methane}
\end{center}
\end{figure}

\begin{figure}[h!]
\begin{center}
\includegraphics[scale=1.2]{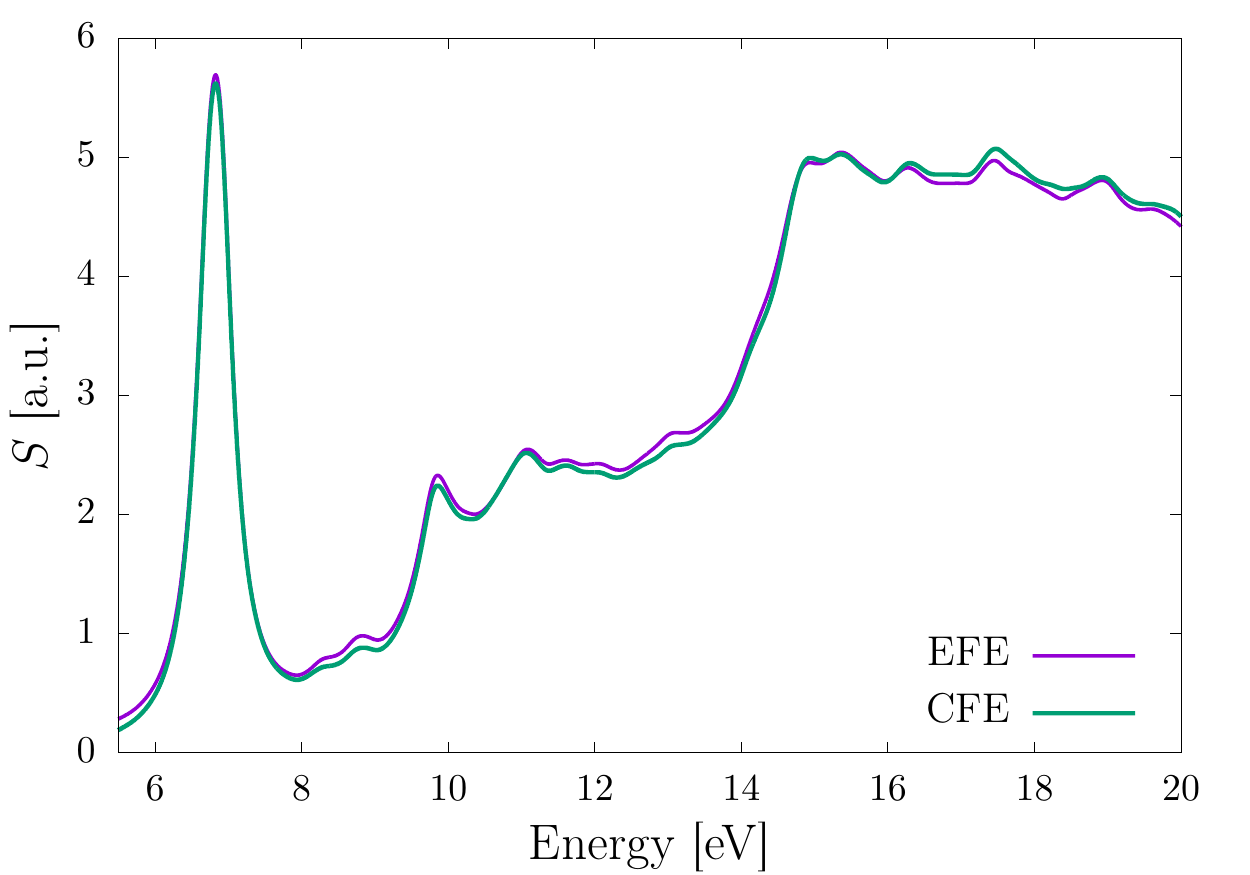}
\caption{\small Absorption spectrum of benzene.}
\label{fig:benzene}
\end{center}
\end{figure}

We now consider methane and benzene molecules as our benchmark systems to compare the performance of the EFE and CFE basis. For both the molecules, we excite the system from their groundstate, in each of the three directions, separately, using a weak Gaussian electric field of the form $\textbf{E}_0(t)=\kappa e^{-(t-t_0)^2/2s^2}\hat{\bx}_j$, with $\kappa=2\times10^{-5}$, $t_0=3$, $s=0.2$ (all in a.u.), and $\hat{\bx}_j$ being the unit vector along $j-$th coordinate axis. Both the systems are propagated for 20 fs. We use a third-order (HEX64SPECTRAL) and fourth-order (HEX125SPECTRAL) finite element mesh for the EFE and CFE based calculations, respectively, adaptively refined as per the strategy presented in Sec.~\ref{sec:efficient}, so as to attain 0.1 mHa accuracy in the groundstate energy per atom. Furthermore, we use a large cubical domain of length 60 a.u.\ and 70 a.u.\ for methane and benzene, respectively, to ensure that the Kohn-Sham orbitals decay to zero, and thereby, avoid any reflection effects. We chose a time-step ($\Delta t)$ of 0.1 and 0.05 a.u.\, for the EFE and CFE calculations, respectively, and a Krylov subspace tolerance ($\epsilon$) of $10^{-7}$ for both the basis sets, to attain $<1$ mHa accuracy in the excitation energies. To evaluate the excitation energies, we first, take the Fourier transform of
the dipole moment to obtain the dynamic polarizability, $\alpha_{i,j}(\omega)$, where $i,j$ are the index of the electric field’s polarization direction and measurement direction of the dipole, respectively. Subsequently, we compute the absorption spectrum $S(\omega)= \frac{4\omega}{3\pi}\frac{\text{Tr}\left[\text{Im}\left[\boldsymbol{\alpha}(\omega)\right]\right]}{\mathlarger{\mathcal{F}}\left[\textbf{E}_0\right](\omega)}$, where $\mathlarger{\mathcal{F}}\left[\textbf{E}_0\right](\omega)$ denotes the Fourier transform of the applied electric field $\textbf{E}_0(t)$. Finally, the peaks in the absorption spectrum correspond to the excitation energies. 
We damp the dipole moment with an exponential window of the form $g(t)=e^{-t/\tau}$, with $\tau=100$ a.u.\ to artificially broaden the excitation peaks. Fig.~\ref{fig:methane} and Fig.~\ref{fig:benzene} compare the EFE and CFE basis based absorption spectrum for methane and benzene, respectively. We attain remarkable agreement in the absorption spectrum between the two basis, underscoring the accuracy of the EFE basis. Table~\ref{tab:compare} compares the performance of the EFE and CFE basis, in terms of the number of basis functions required and the computational cost incurred. As evident, the EFE basis attains a staggering $\sim50\times$ and $\sim100\times$ speedup over the CFE basis for the methane and the benzene molecule, respectively. This speedup is attributed to a combination of $3-5\times$ reduction in number of basis functions, $\sim2\times$ reduction in the stencil of the discrete matrices (HEX64SPECTRAL in EFE and HEX125SPECTRAL in CFE), $2\times$ reduction in the time-step, and $4-5\times$ reduction in Krylov subspace size required by the EFE as compared to that of the CFE basis. Given the huge computational cost associated with the CFE basis, in the remaining benchmark calculations presented in subsequent sections, we only employ the EFE basis.

\begin{table}[h!]
    \caption{\small Comparison of EFE and CFE basis sets: number of basis function ($M$); the minimum ($h_{\text{min}}$) and maximum ($h_{\text{max}}$) FE mesh size;  total computational resources required in CPU node hours for the enitre simulation of 20 fs ($T_C$). For the EFE basis, the value in parenthesis specifies the number of enrichment functions.} 
\begin{tabular}{|p{1.4cm} | c | c | c | c | c | c |}
\hline
\multirow{2}{1.4cm}{System} & 
\multicolumn{2}{c|}{$M$} & \multicolumn{2}{c|}{$h_{\text{min}}$, $h_{\text{max}}$}  & \multicolumn{2}{c|}{$T_C$}   \\
\cline{2-7}
& CFE & EFE & CFE & EFE & CFE & EFE\\
\hline \hline
methane & 427,721 & 81,634 (34) & 0.04, 3.97 & 0.26, 3.97 & 556 & 12 \\
\hline
benzene & 1,261,149 & 427,721 (114) & 0.04, 3.52 & 0.26, 3.52 & 6,419 & 66\\
\hline
\end{tabular}
\label{tab:compare}
\end{table}

\subsection{Weak perturbation on sodium nanoclusters} \label{sec:NaCluster}
\begin{figure}[h!]
\begin{center}
\includegraphics[scale=1.2]{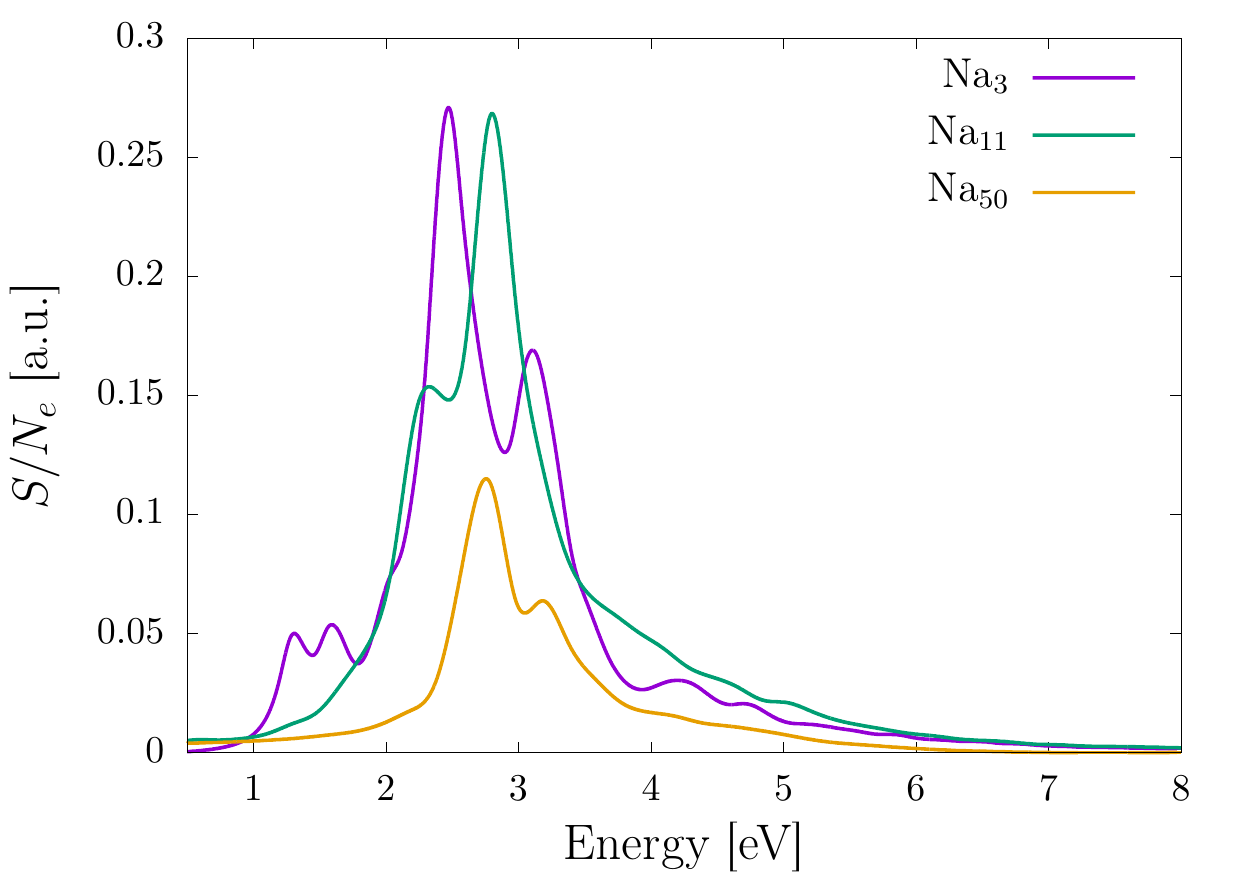}
\caption{\small Absorption spectrum of various sodium cluster, normalized with number of electrons ($N_e$).}
\label{fig:NaCluster}
\end{center}
\end{figure}

In order to demonstrate the efficacy of the EFE for larger systems, we study the absorption spectrum of three sodium clusters of increasing sizes---Na$_3$, Na$_{11}$, and Na$_{50}$. For all the three systems, we employ a third-order (HEX64SPECTRAL) finite element mesh that is adaptively refined using the strategy presented in Sec.~\ref{sec:efficient} and commensurate with 0.1 mHa accuracy in the groundstate energy per atom. Furthermore, we use large cubical domains---70 a.u.\ for Na$_3$ and Na$_{11}$, and 85 a.u.\ for Na$_{50}$---to avoid any reflection effects from the boundary. We use the same Gaussian electric field as used in the methane and benzene benchmark calculations to excite the clusters from their groundstate and propagate the TDKS equations for 20 fs. Similar to the previous examples, we use a time-step of 0.1 a.u.\ and Krylov subspace tolerance of $10^{-7}$ to attain $<1$ mHa accuracy in excitation energies. Fig.~\ref{fig:NaCluster} presents the absorption spectrum of the three sodium clusters, normalized with the number of electrons, wherein we have used an exponential window of the form $g(t)=e^{-t/\tau}$, with $\tau=200$ a.u.\, to artificially broaden the excitation peaks. Table~\ref{tab:NaCluster} provides the number of basis functions and the total computation time required by the EFE basis for the three sodium clusters. Using the timings reported, we attain a scaling of $\mathcal{O}(N_e^{2.01})$ with respect to the number of electrons, which agrees well with the expected quadratic scaling of the TDKS problem.  

\begin{table}[h!]
    \caption{\small Comparison of number of basis functions ($M$) and total computational resources in CPU node hours for the entire simulation of 20 fs ($T_C$), as required by the EFE basis for various sodium clusters.} 
\begin{tabular}{M{0.1\columnwidth}M{0.15\columnwidth}M{0.15\columnwidth}}
\hline 
\hline
    System & $M$ & $T_C$\\ \hline
    Na$_3$ & 120,886 & 17 \\
    Na$_{11}$ & 446,866 & 153\\
    Na$_{50}$ & 1,734,355 & 3,486\\
    \hline
\end{tabular}
\label{tab:NaCluster}
\end{table}

\subsection{Weak to strong perturbations on green fluorescent protein}
\begin{figure}[t!]
    \centering
    \begin{subfigure}{0.45\columnwidth}
        \centering
        \includegraphics[scale=0.6]{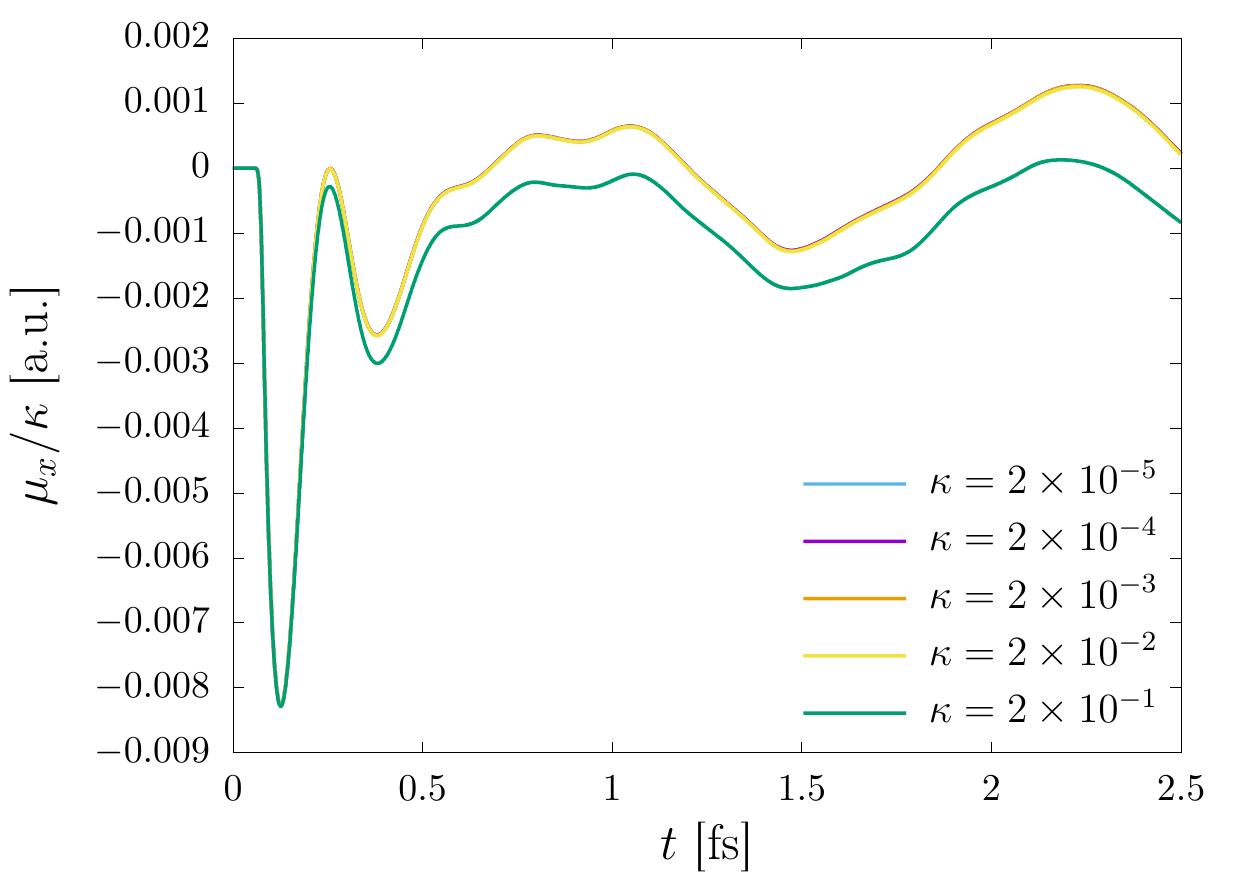}
        \caption{}
        \label{fig:gfp_dipole}
    \end{subfigure}%
    \begin{subfigure}{0.45\textwidth}
        \centering
        \includegraphics[scale=0.6]{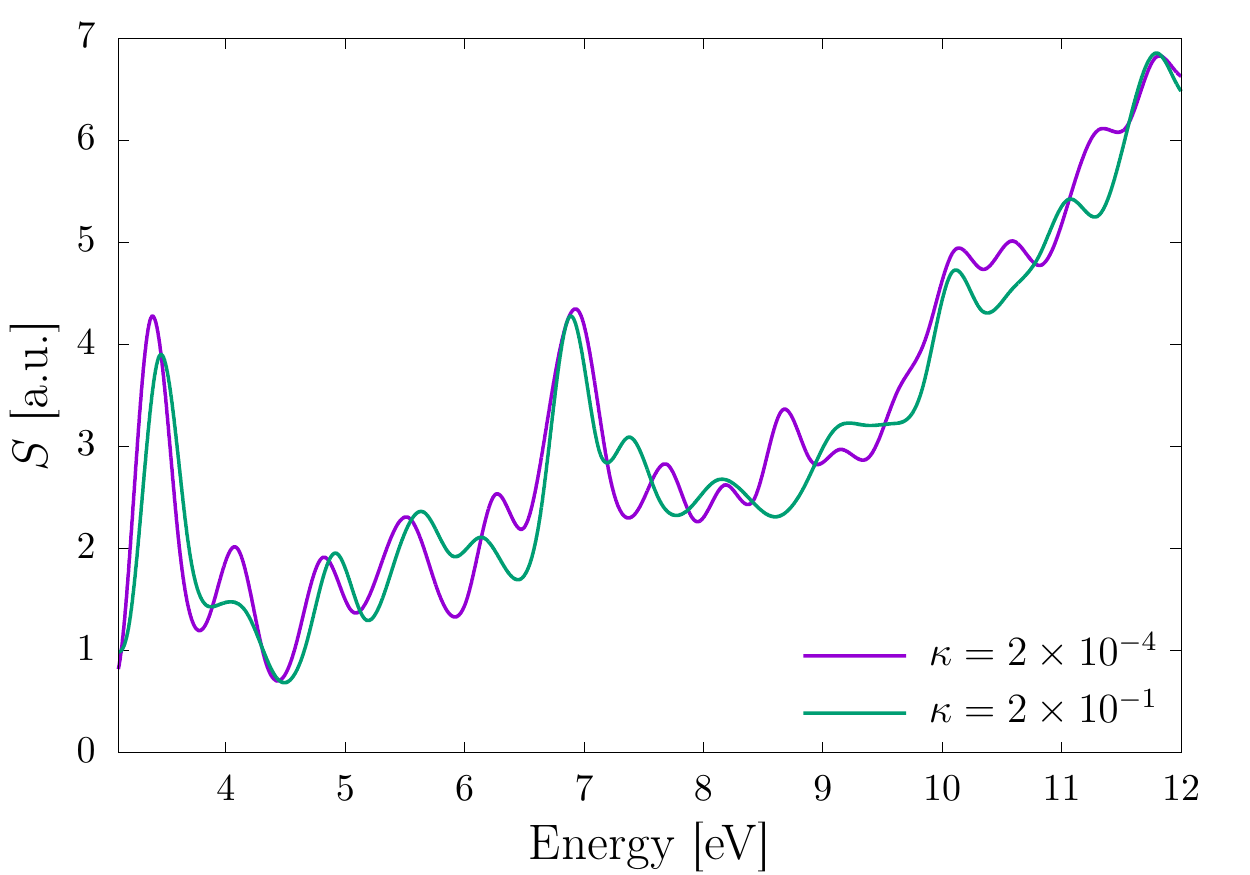}
        \caption{}
        \label{fig:gfp_absorption}
    \end{subfigure}
    \label{fig:gfp}
    \caption{Comparison of dipole moment and absorption spectrum of the GFP choromophore for different strengths of the perturbing electric field. (a) Comparison of the scaled dipole moment at various values of the electric field strength ($\kappa$). (b) Comparison of the absorption spectrum for weak ($\kappa=2\times10^{-4}$) and strong ($\kappa=2\times10^{-1}$) perturbation.}
\end{figure}

The green fluorescent protein (GFP), abundantly found in jellyfish, corals, copepods, is a commonly used tool to understand a wide array of biochemical processes ranging from fluorescence microscopy~\cite{Bastiaens2000} to gene therapy~\cite{Wahlfors2001} to reporter gene technology~\cite{Naylor1999}. In this example, we study the linear to nonlinear response of the chromophore of the GFP---a small molecule that lends it color. We employ a second-order (HEX27) finite element mesh that is adaptively refined using the strategy presented in Sec.~\ref{sec:efficient} and commensurate with 0.1 mHa accuracy in the groundstate energy per atom. We use a large cubucal domain of 120 a.u.\ to avoid any reflection effects from the boundary. First, we identify the onset of nonlinear response by subjecting the chromophore, at its groundstate, to a series of Gaussian electric field of the form provided in Sec.~\ref{sec:methane_benzene} with the strength ($\kappa$) taking values of $2\times10^{-5}$, $2\times10^{-4}$, $2\times10^{-3}$, $2\times10^{-2}$, and $2\times10^{-1}$ (all in a.u.). In linear response regime, the dipole moments scaled by the strength of the electric field are expected to be identical, and hence, can be used to identify the onset of nonlinear response. For the purpose of identifying the linear to nonlinear transition, for each of the strengths, we propagate the system for a short duration of 2.5 fs and observe the scaled dipole moment scaled. As evident from ~\ref{fig:gfp_dipole}, the clear deviation of the scaled dipole moment for $\kappa=2\times10^{-1}$ from that of the rest of the strengths, indicates the onset of nonlinear regime to be around $\kappa=2\times10^{-1}$. Having identified the linear and nonlinear regimes, we propagate the system for longer timescale of 20 fs using a weak ($\kappa=2\times10^{-4}$) and a strong perturbation ($\kappa=2\times10^{-1}$). Fig.~\ref{fig:gfp_absorption} presents a comparison of the absorption spectrum for both the strengths, where the peaks are artificially broadened with an exponential window $e^{-t/\tau}$ with $\tau=300$ a.u.~. We observe the first excitation peak at $3.37$ eV, which agrees well with the experimental value of 3.51 eV~\cite{Dong2006}.  As evident, the strong perturbation has a decreased height of the first absorption peak, largely owing to saturation effect (i.e., the inability of the highly excited molecule under strong perturbation to absorb subsequent radiations). Furthermore, for the strong perturbation, we observe a blue-shifting of the absorption peaks in the lower energy ($< 6$ eV) regime. In the higher energy regime ($> 6$ eV), for the strong perturbation, we observe fewer excitation peaks, indicating the absorption to be spread over a range of frequencies. We remark that, despite the use of strong perturbation and large domain size, we require only $\sim18,000$ basis functions per atom, highlighting the efficiency of the EFE basis even for strong perturbations in TDDFT.     

\subsection{Strong perturbation on Mg$_2$} \label{sec:Mg2}
\begin{figure}[h!]
\begin{center}
\includegraphics[scale=1.2]{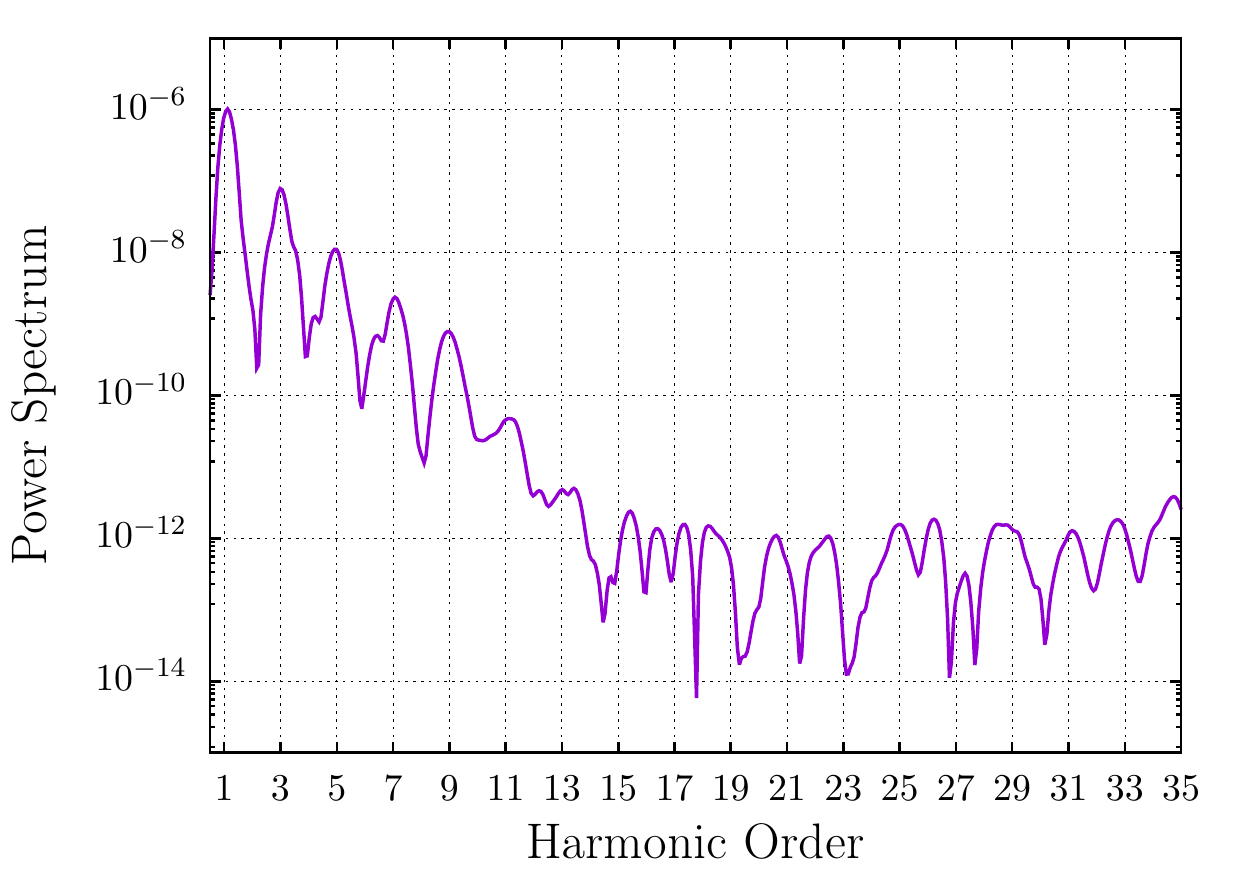}
\caption{\small Power spectrum of Mg$_2$.}
\label{fig:Mg2}
\end{center}
\end{figure}

We examine the competence of the EFE basis for simulating strong perturbations by studying the higher harmonic generation in a magnesium dimer with a bond-length of $4.74$ a.u.~. We use a strong sinusoidal electric field of the form $\textbf{E}_0(t)=\kappa \text{sin}(\pi t/T)\text{sin}(\omega t)\hat{\bx}_1$, with $\kappa=0.01$, $\omega=0.03$, and $T= 5\times(2\pi/\omega)$ (all in a.u.) to excite the system from its groundstate and propagate for $T$ fs. For an efficient choice of finite element mesh, we employ adaptive third-order (HEX64SPECTRAL) finite elements, obtained using the approach presented in Sec.~\ref{sec:efficient}. As with previous examples, we use a cubical domain of length 100 a.u.\ to eliminate any reflection effects from the boundaries. We employ a time-step of 0.1 a.u.\ and a Krylov subspace tolerance of $10^{-7}$. Fig.~\ref{fig:Mg2} shows the dipole power spectrum ($P(\omega)$) for Mg$_2$, defined as $P(\omega)=\modulus{\int_0^Te^{-i\omega t}\frac{d}{dt^2}\mu_x(t)dt}^2$ (i.e., square of the Fourier transform of the acceleration of the dipole moment). We have artificially broadened the peaks by using a gaussian window of form $e^{-t^2/\tau}$, with $\tau=10^5$ a.u.~. For a system with spatial inversion symmetry, only odd multiples of the frequency of the exciting laser pulse must be emitted, which is well verified in Fig.~\ref{fig:Mg2} with the peaks in the power spectrum coinciding with odd harmonics (odd multiples of $\omega$). We also observe that the rate of decay of the intensity of the peaks flattens beyond the $13$-$th$ harmonic, which corroborates well with the plateau phenomenon observed in experiments~\cite{Brabec2000}. Despite the large domain size used in this calculation, we require only $\sim 32,000$ basis functions per atom, further underscoring the efficiency of the EFE basis even for strong perturbations in TDDFT.

\subsection{Parallel Scalability}
\begin{figure}[h!]
\begin{center}
\includegraphics[scale=1.2]{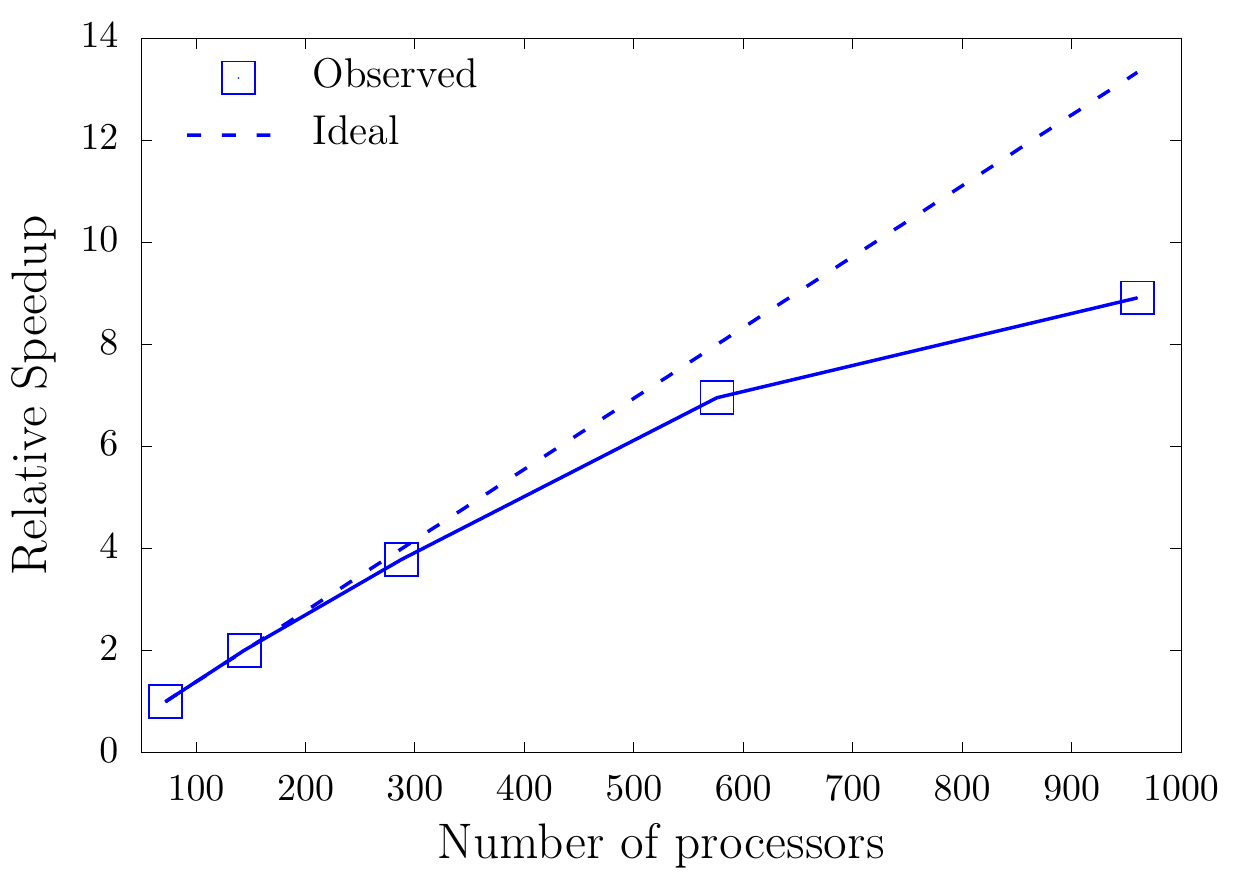}
\caption{\small Parallel scalability of the EFE implementation.}
\label{fig:scalability}
\end{center}
\end{figure}

Finally, we examine the parallel scalability of the implementation of our EFE basis for TDDFT calculations, using the Na$_{50}$ cluster with $~\sim1.7$ million basis functions as a benchmark system. We repeat the same calculation as presented in Sec.~\ref{sec:NaCluster} and measure the relative speedup in the walltime with respect to 72 processors. As evident from Fig.~\ref{fig:scalability}, we attain close to ideal scaling until 576 processors, at which we observe a parallel efficiency of 87\%. The efficiency drops to 67\% at 960 processors owing to fact that,
at 960 processors, the number of basis functions per processor falls below 2000, which is low to sustain good parallel scalability.

\section{Conclusion} \label{sec:conclusion}
We have presented an efficient mixed basis, termed as enriched finite element (EFE) basis, for all-electron real-time TDDFT calculations, constructed by augmenting the classical finite element (CFE) basis with enrichment functions obtained from single atom groundstate orbitals and electrostatic potentials. In effect, the EFE basis combines the efficiency of the atomic orbitals, which capture part of the physics, with the completeness of the CFE basis, which in turn, guarantees systematic convergence. In particular, we orthogonalized the enrichment functions with respect to the underlying CFE basis to ensure a well-conditioned basis. Additionally, we employed the knowledge of groundstate electronic fields to obtain an efficient adaptive finite element mesh. We established close to optimal rate of convergence in the dipole moment with respect to both spatial and temporal discretization. Notably, we demonstrated a striking $50-100\times$ speedup afforded by the EFE basis over the CFE basis, while being commensurate with the desired chemical accuracy. We assessed the performance of the EFE basis for handling large systems by studying the absorption spectrum of sodium clusters of increasing sizes. Furthermore, for the sodium clusters considered, we attain an almost quadratic computational complexity with respect to number of electrons, as is theoretically expected. Importantly, we also established the efficacy of the EFE basis for strong perturbations and the accompanying requirement of large domain sizes by examining transition from linear to nonlinear response in the GFP chromophore as well as by studying the higher harmonic generation in Mg$_2$. Finally, we obtained good parallel scalability up to $\sim1000$ processors for a benchmark Na$_{50}$ system containing $\sim1.7$ million basis functions. 

Overall, the EFE basis offers a robust, efficient, systematically convergent, and scalable basis for all-electron TDDFT calculations. Given the importance of relativistic effects
in all-electron calculations, our future work will involve an extension of the proposed EFE basis to incorporate both scalar relativistic and spin-orbit coupling effects. Further, we intend to combine the configurational forces formulation for EFE basis~\cite{Rufus2022} to enable efficient all-electron TDDFT based Ehrenfest dynamics calculations. Lastly, the EFE basis, paves the way for an efficient and systematic route to study the transferability of widely used pseudopotentials for electron dynamics.


\section*{Acknowledgements}
We thank the support from the Department of Energy, Office of Basic Energy Sciences, Grant No. DE-SC0017380, under the auspices of which the computational framework for the enriched finite element basis was developed. This research used resources of the National Energy Research Scientific Computing Center, a DOE Office of Science User Facility supported by the Office of Science of the U.S. Department of Energy under Contract No. DE-AC02-05CH11231. This work used the Extreme Science and Engineering Discovery Environment (XSEDE), which is supported by National Science Foundation Grant number ACI-1053575. V.G. also acknowledges the support of the Army Research Office through the DURIP Grant No. W911NF1810242, which also provided the computational resources for this work.

\bibliography{ref}

\end{document}